\documentclass[preprint,5p,times,twocolumn]{elsarticle}
\usepackage{lineno,hyperref}
\usepackage{graphicx}
\usepackage[font=normalfont,labelfont=bf]{caption}
\usepackage{lipsum}
\usepackage{amssymb}
\usepackage{graphicx}
\usepackage{xcolor}
\usepackage[export]{adjustbox}
\usepackage{float}
\usepackage{caption}
\graphicspath{ {./images/} }

\journal{Astronomy and Computing}

\begin{document}

\begin{frontmatter}

\title{Rosetta: a container-centric science platform for\\ resource-intensive, interactive data analysis.}

\author[INAF,UNITS]{Stefano Alberto Russo\corref{cor1}}
\ead{stefano.russo@inaf.it}
\cortext[cor1]{Corresponding author}
\author[INAF]{Sara Bertocco}
\author[IRA]{Claudio Gheller}
\author[INAF]{Giuliano Taffoni}

\address[INAF]{INAF (Italian National Institute for Astrophysics) – Astronomical Observatory of Trieste, Italy}
\address[IRA]{INAF (Italian National Institute for Astrophysics) – Institute of Radioastronomy, Bologna, Italy}
\address[UNITS]{Department of Mathematics and Geosciences, University of Trieste, Italy}

\begin{abstract}
Rosetta is a science platform for resource-intensive, interactive data analysis which runs user tasks as software containers.
It is built on top of a novel architecture based on framing user tasks as microservices - independent and self-contained units - which allows to fully support custom and user-defined software packages, libraries and environments.
These include complete remote desktop and GUI applications, besides common analysis environments as the Jupyter Notebooks.
Rosetta relies on Open Container Initiative containers, which allow for safe, effective and reproducible code execution; can use a number of container engines and runtimes; and seamlessly supports several workload management systems, thus enabling containerized workloads on a wide range of computing resources.
Although developed in the astronomy and astrophysics space, Rosetta can virtually support any science and technology domain where resource-intensive, interactive data analysis is required.
\end{abstract}


\end{frontmatter}

%%======================================
%%        Introduction
%%======================================
\section{Introduction}
\label{sec:introduction}

Data volumes are rapidly increasing in several research fields, as in bioinformatics, particle physics, earth sciences, and more. Next generation sequencing technologies, new particle detectors, recent advances in remote sensing techniques and higher resolutions in general, on both the instrumental and the simulation side, are constantly setting new challenges for data storage, processing and analysis.

Astrophysics is no different, and the upcoming generation of surveys and scientific instruments as the Square Kilometer Array (SKA) \citep{SKA}, the Cherenkov Telescope Array (CTA) \citep{CTA}, the Extremely Large Telescope (ELT) \citep{EELT}, the James Webb Space telescope \citep{jwst}, the Euclid satellite \citep{Euclid} and the eROSITA All-Sky Survey \citep{erosita} will pile up on this trend, bringing the data volumes in the exabyte-scale. Moreover, numerical simulations, a theoretical counterpart capable of reproducing the formation and evolution of the cosmic structures of the Universe, must reach both larger volumes and higher resolutions to cope with the large amount of data produced by current and upcoming surveys. State of the art cosmological N-body hydrodynamic codes (as OpenGADGET, GADGET4 \citep{gadget4} and RAMSES \citep{ramses}) can generate up to 20 petabytes of data out of a single simulation run, which are required to be further post-processed and compared with observational data \citep{2018MNRAS.475..676S,2017A&C....20...52R,2019ASPC..521..567T,2016NewA...42...49H}.

The size and complexity of these new experiments (both observational and numerical) require therefore considerable storage and computing resources for their data to be processed and analyzed, and possibly to adopt new approaches and architectures. 
High Performance Computing (HPC) systems including Graphical Processing Units (GPUs) and Field Programmable Gate Arrays (FPGAs), together with the so called ``bring computing close to the data" paradigm are thus becoming key players in obtaining new scientific results \citep{asch2018}, not only by reducing the time-to-solution, but also by becoming the sole approach capable of processing datasets of the expected size and complexity.

In particular, even the last steps of the data analysis processes, which could be usually performed on researchers' workstations and laptops, are getting too resource-intensive and progressively required to be offloaded to such systems as well.

Although capable of satisfying the necessary computing and storage requirements, these systems are usually hosted in remote computing centers and managed with queue systems, in order to dynamically share their resources across different users and to optimize the workload and the throughput. This can strongly complicate the user interaction, requiring remote connections for shell and graphical access (as SSH and X protocol forwarding), careful data transfer and management, and scheduler-based access to computing resources which strongly limits interactive access to the system.
Bringing along the software required for the analysis can be even more challenging, and without proper setup (in particular with respect to its dependencies) it can not only fail to start or even compile, but also severe reproducibility issues can arise \citep{bhandari2019characterization}.

To address these challenges, we see an increasing effort in developing the so called \emph{science platforms} \citep{2020ASPC..527..777T,sciplat1,sciplat2,sciserv}. A science platform (SP) is an environment designed to offer users a smoother experience when interacting with remote computing and storage resources, in order to mitigate some of the issues outlined above. %in the previous paragraph.

In science and, more specifically, in astronomy, a number of SPs have been designed and developed over the past years. 

CERN SWAN \citep{piparo2018swan} represents CERN's effort to build towards the science platform paradigm. SWAN is a service for interactive, web-based data analysis which makes Jupyter Notebooks widely available on CERN computing infrastructure together with a Dropbox-like solution for data management. However, as of today, this solution does not provide support for applications other than the Jupyter Notebooks and a built-in shell terminal, does not allow using custom or graphical software environments and requires heavy system-level integration in order to be used on top of existent computing resources.

ESA Datalabs \citep{datalabs} is a science platform specific to astronomy and astrophysics. Similarly to CERN SWAN, it allows users to work on ESA's computing infrastructure using interactive computing environments as Jupyter Lab and Octave (or to choose from pre-packaged applications as TOPCAT). Datalabs is mainly focused on enabling users to gain direct access to ESA's datasets, it does not support using custom software environments, and it is not an open source project.

The Large Synoptic Survey Telescope (LSST) developed a similar science platform \citep{juric2017lsst}, based on a set of integrated web applications and services through which the scientific community will be able to ``access, visualize, subset and analyze LSST data". The platform vision document does not mention applications other than the Jupyter Notebooks, nor support for custom or graphical software environments, and refers to its own computing architecture. 

There are also a number of initiatives entirely focus on supporting Jupyter Notebooks on cloud and HPC infrastructures (such as \citep{hpcnotebooks}, \citep{metabolomicsjupyter}, \citep{milliganjupyter} and \citep{castronova2018general}), which might fall in our SP definition to some extent, and in particular in Astronomy and Astrophysics it is worth to mention SciServer \citep{sciserv}, Jovial \citep{jovial} and CADC Arcade \citep{canfar}. 

Lastly, it has to be noted that the private sector is moving fast with respect to resource-intensive and interactive data analysis, mainly driven by the recent advances in artificial intelligence and machine learning. In this context, we want to cite Google Colab \citep{bisong2019google} and Kaggle Notebooks \citep{kagglenotebooks}, which are built around heavily customised versions of the Jupyter Notebooks, and Azure Machine Learning \citep{azureml}, which provides a nearly full-optional SP specifically targeted at machine learning workflows. 

While on one hand all of the above mentioned SPs do make it easier to access and use remote computing resources, on the other, since they are mainly focused on web-based and integrated analysis environments built on top Jupyter Notebooks or similar software, they also introduce two main drawbacks:

\begin{enumerate}
    \item users are restricted in using pre-defined software packages, libraries and environments, which besides constraining their work can also lead to reproducibility issues, and
    \item graphical software environments as remote desktops and GUI applications are supported only to a limited extent, if not completely unsupported.
    
\end{enumerate}

Moreover, the deployment options for most of the SPs developed today rely on technologies originating from the IT industry (e.g. Kubernetes) and require deep integration at system-level, which is often hard to achieve in the framework of HPC clusters and data-intensive system. This is not only because of technological factors and legacy aspects, but also because of a generalized pushback for exogenous technologies from some parts of the HPC community \citep{nih_hpc,nih2_hpc,nih3_hpc,nih4_hpc}.

In this paper we present a science platform which aims at overcoming these limitations: \textit{Rosetta}. Built on top of a novel architecture based on framing user tasks as microservices - independent and self-contained units - Rosetta allows to fully support custom software packages, libraries and environments, including remote desktops and GUI applications, besides standard web-based analysis environments as the Jupyter Notebooks. Its user tasks are implemented as software containers \citep{suse_cont}, which allow for safe, effective and reproducible code execution \citep{boettiger2015introduction}, and that in turn allows users to add and use their own software containers on the platform.

Rosetta is also designed with real-world deployment scenarios in mind, and thus to easily integrate with existing computing and storage resources including HPC clusters and data-intensive systems, even when they do not natively support containerization.

Although astronomy remains its mainstay (Rosetta has been developed in the framework of the EU funded project ESCAPE\footnote{ESCAPE aims to address the open science challenges shared by SKA, CTA, KM3Net, EST, ELT, HL-LHC, FAIR as well as other pan-European research infrastructures as CERN, ESO, JIVE in astronomy and particle physics.}), Rosetta can virtually support any science and technology domain.

This paper is organized as follows. In Sections \ref{sec:architecture}, \ref{sec:implementation} and \ref{sec:security}, we discuss the architecture of the Rosetta platform, its implementation and the security aspects. 
This is followed, in Section~\ref{sec:rosetta}, by an overview of the platform from a user prospective.
Next, we present the deployment and usage scenario in a real production environment and a few use cases we are supporting (Section~\ref{sec:usecases}), leaving the last section to conclusions and future work.

%%======================================
%%   Architecture
%%======================================
\section{Architecture}
\label{sec:architecture}

Rosetta's architecture is entirely designed to provide simplified access to remote, dynamically allocated computing and storage resources without restricting users to a set of pre-defined software packages, libraries and environments.
It unfolds in two main components: the \textit{platform architecture} and the \textit{task orchestration architecture}.

\begin{figure}[htpb]
\centering \includegraphics[scale=0.18]{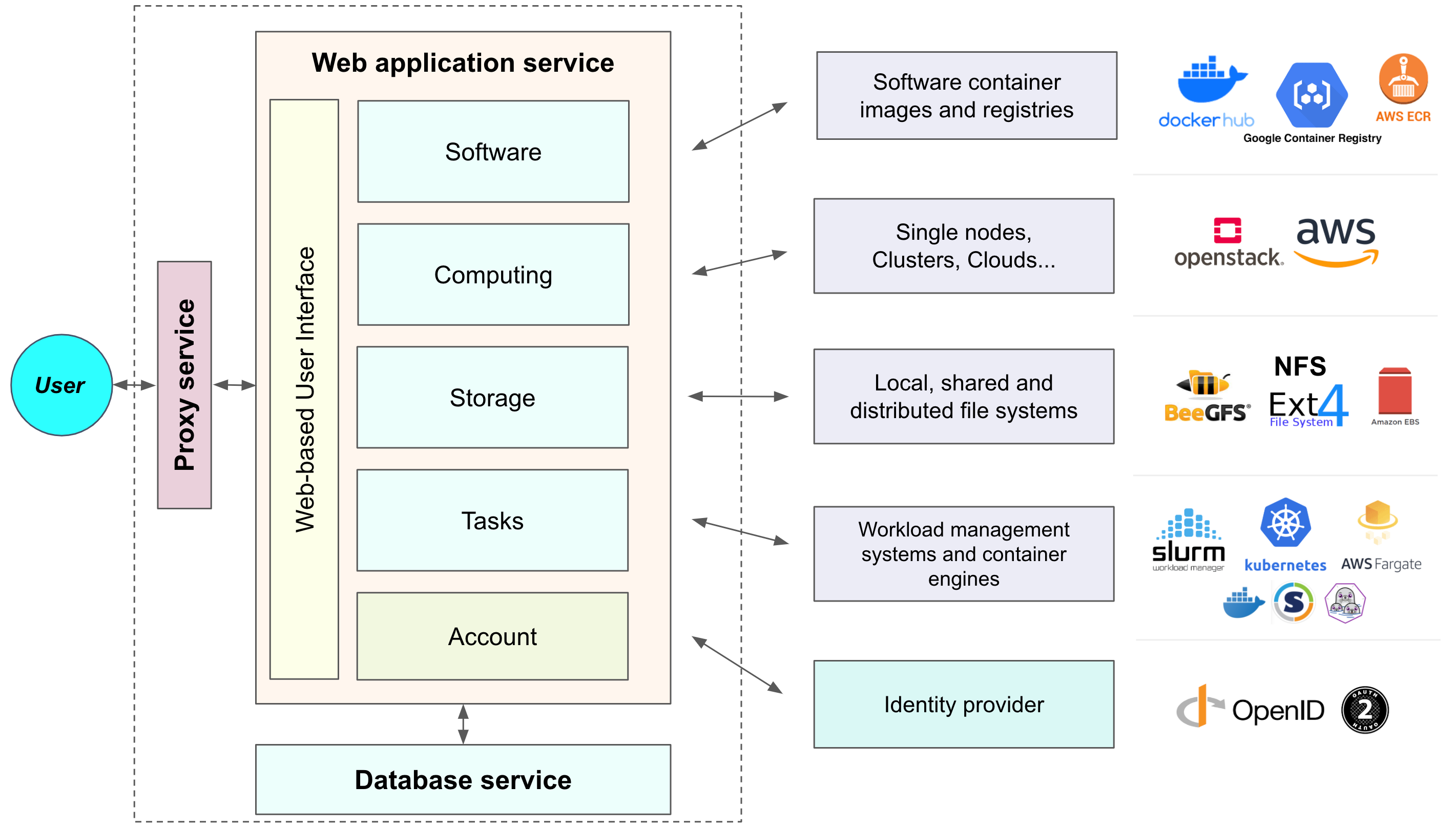}
\caption{Rosetta main architecture. The first level of abstraction consist in the proxy, database and web application services. The web application service is further break down into its main components (software, computing, storage, tasks and account) together with their real world counterparts, some examples of which are given in the right part of the figure.} 
\label{fig:arch}
\end{figure}

The platform architecture follows a standard approach where a set of services implement the various functionalities, and it is schematized in Figure~\ref{fig:arch}. These comprise a web application service for the main application logic and the web-based UI, a database service for storing internal data and a proxy service for securing the connections. The web application service functionalities can be further grouped in modules which are responsible for managing the software containers, interacting with the computing and storage resources, orchestrating the user tasks, handling the user authentication and so on. 

In particular:
\begin{itemize}
    \item  \emph{Software} functionalities allow to track  the software containers available on the platform, their settings and container registries\footnote{A container registry is a place where container images are stored, which can be public or private, and deployed both on premises or in the Cloud. Many container registries can co-exisist at the same time.};
    \item \emph{Computing} functionalities allow to interact with both standalone and clustered computing resources, hosted either on premises (e.g. via Openstack) or on cloud systems (e.g. on Amazon AWS);
    \item \emph{Storage} functionalities allow browsing and operating on local and shared file system (as Ext4, NFS, BeeGFS); 
    \item \emph{Task} functionalities allow submitting and stopping tasks as well as viewing their logs, by interacting with the computing resources workload management systems (WMSs) as Slurm and Kubernetes and/or their container engines (e.g. Docker, Singularity, Podman);
    \item \emph{Account} functionalities provide user account and profile management features including user registration, login and logout, supporting both local and external authentication (e.g. OpenID Connect, Shibbolet).
\end{itemize}

Rosetta's task orchestration architecture follows instead  a novel, microservice-oriented architecture \citep{O1-124_adassxxx} based on software containers.
Microservices \citep{newman2015building} are independent, self-contained and self-consistent units that perform a given task, which can range from a simple functionality (e.g. serving a file to download) to complex computer programs (e.g. classifying images using a neural network). They are interacted with using a specific interface, usually a REST API over HTTP, which is exposed on a given port.
Microservices fit naturally in the containerisation approach, where each microservice runs in its own container, isolated from the underlying operating system, network, and storage layers. 
User tasks in Rosetta are thus always executed as software containers, and treated as microservices. Rosetta can therefore stay agnostic with respect to the task interface, some examples of which include a Jupyter Notebook server, a web-based remote desktop or a virtual network computing (VNC) server, but also a secure shell (SSH) server with X protocol forwarding is a perfectly viable choice.

One of the main features of this approach, where user tasks are completely decoupled from the platform, is to make it possible for the users to add their own software containers. There is indeed no difference between ``platform'' and ``user'' containers, as long as they behave as a microservice. Rosetta users can thus upload their own software containers on a container registry, add them in the platform by setting up a few parameters (as the container image and the interface port), and then use them for their tasks.

In order to make use of this architecture for user tasks orchestration, Rosetta needs to be able to submit to the computing resources a container for execution, and to know how to reach it (i.e. on which IP address).
These functionalities are standard and built-in in most modern container orchestrators (e.g Kubernetes), however as mentioned in the introduction Rosetta has been designed to also support computing resources not natively supporting containerized workloads (e.g. HPC clusters and data-intensive systems). On these computing resources, also depending on the WMS and container engine used, some key features might not be available, as full container-host filesystem isolation, network virtualization and TCP/IP traffic routing between the containers.
To work around these missing features, Rosetta relies on an \emph{agent}, which is a small software component in charge of helping to manage the task container life cycle. Its main features comprises setting up the environment for the container execution, managing dynamic port allocation, reporting the host IP address to the platform, and running the container itself. The agent internal logic is described more in detail in section \ref{subsec:tasks}.

When a container is started, its interface has to be made accessible by the user. This is achieved first by making the interface port reachable on the internal network between the computing resource and Rosetta, and then by exposing it to the outside world through Rosetta itself, thus making it accessible by the user.
The first step can make use of simple TCP/IP tunnels as well as more sophisticated techniques usually available in modern container orchestrators and WMSs, while the second one can be accomplished either by directly exposing the task interface as-is or by relaying on a proxy service, which also allows to enforce access control and connection encryption.

\begin{figure}[htbp]
\centering\includegraphics[scale=0.20]{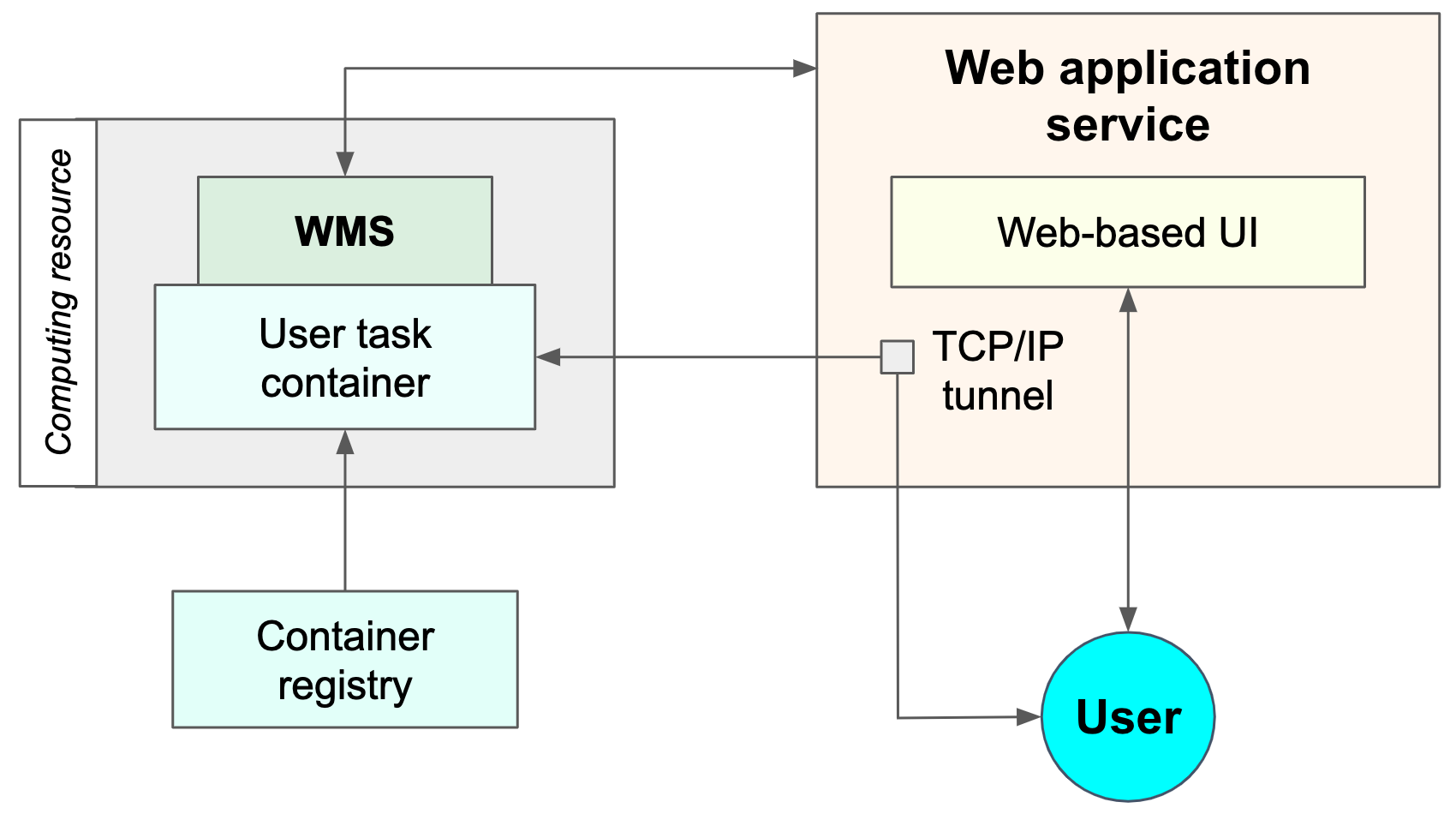}
\caption{Rosetta user task orchestration using the computing resource's WMS and a direct connection to the task interface through a TCP/IP tunnel.}
\label{fig:task_wms_direct}
\end{figure}

\begin{figure}[htbp]
\centering\includegraphics[scale=0.20]{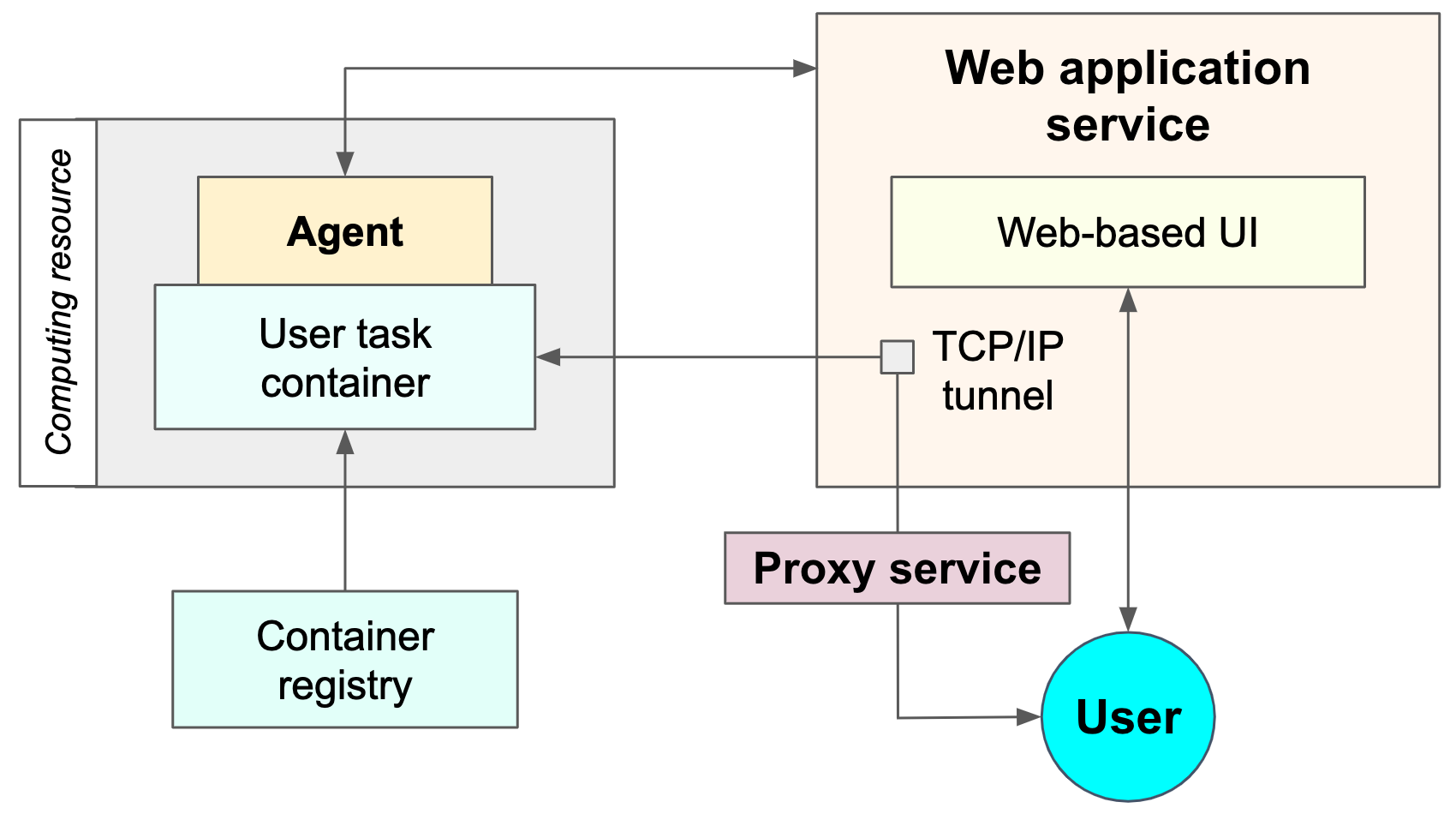}
\caption{Rosetta user task orchestration using the agent and the proxy service on top of a TCP/IP tunnel for connecting to the task interface.}
\label{fig:task_agent_proxy}
\end{figure}

Once tasks are executed and their interfaces made accessible, no further operations are required, and the users can be looped in.
A diagram of this flow is presented with two examples: the first using a WMS supporting containerized workloads with direct connection to the task interface (Figure~\ref{fig:task_wms_direct}), the second using the agent to run the task container and relaying on the proxy for connecting to the task interface (Figure~\ref{fig:task_agent_proxy}).

%%======================================
%%    Implementation
%%======================================
\section{Implementation}
\label{sec:implementation}

Rosetta is entirely built using open-source technologies, in particular Python and the Django web framework, and released as an open source project\footnote{\url{https://www.ict.inaf.it/gitlab/exact/Rosetta}}.
Other technologies include HTML and JavaScript for the UI, Postgres for the database\footnote{The database service can be replaced by any other database supported by Django.} and Apache for the proxy. The platform services (not to be confused with the user tasks software containers) are containerised using the Docker engine and using Docker Compose as the default orchestrator\footnote{Other orchestrators can be  supported as well, e.g. Kubernetes.}. Besides the web application, database and proxy services, Rosetta includes an optional container registry service, which can be used to store software containers locally, and a test Slurm cluster service for testing and debugging.
Rosetta deployment tools provide a set of management scripts to build, bootstrap and operate the platform and a logging system capable of handling both user-generated and system errors, exceptions and stack traces.

\begin{figure}[htpb]
\includegraphics[scale=0.25]{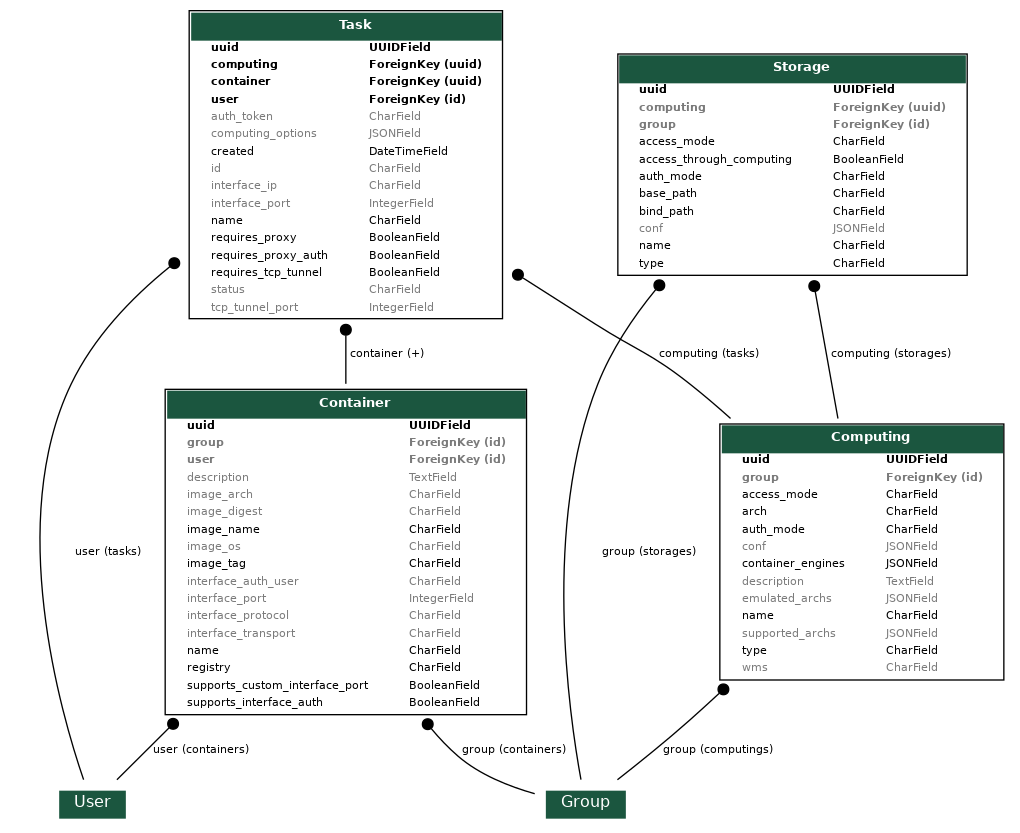}
\caption{The Rosetta Django ORM schema, showing the various models and their relationships. Some minor and less relevant models as the user profile, the login tokens and the key pairs have been excluded for the sake of simplicity.}
\label{fig:ORM}
\end{figure}

The web application functionalities are handled with a combination of Django object–relational mapping (ORM) models and standard Python functions and classes. 
The ORM schema, which represents how the ORM models are actually stored in the database, is summarized in Figure \ref{fig:ORM}. In the following subsections we will describe their implementation according to the grouping introduced in section \ref{sec:architecture}: Software, Computing, Storage, Tasks and Account.

%%======================================
%%  Implementation - Containers
%%======================================
\subsection{Software}
\label{subsec:containers}

Software lives in Rosetta only as software containers. Software containers are represented using a Django ORM model which acts as a twin of the ``real'' container, providing metadata about the container itself. Rosetta relies on Open Container Initiative (OCI) containers, which must be stored on an OCI-compliant container registry.

The \verb`Container` ORM model has a \verb`name` and a \verb`description` fields to represent the container on Rosetta, and a series of attributes to identify its image: the \verb`registry` (to set on which container registry it is hosted), the \verb`image_name` (to locate it on the registry) and the \verb`image_tag` (to set a specific container version). 

The \verb`image_arch`, \verb`image_os` and \verb`image_digest` attributes provide instead more fine-grained control in order to uniquely identify the image, and should be used in production environments. A container image is indeed uniquely identified on an OCI registry only if using, besides its name, either a triplet of tag, architecture and OS or an image hash digest (usually generated with SHA-256). This is because on OCI registries, multiple images can be assigned to the same tag, in order to enable multi-OS and multi-architecture support. Moreover, it has also to be noted that while a tag can be re-assigned, a digest is an immutable identifier and ensures reproducibility.

Containers can be registered in Rosetta as platform  containers or user containers. A platform container is not associated with a specific user and thus available for all of them, while a user container belongs to and is accessible by a specif user only, according to its \verb`user` attribute. Containers can also be shared within (and made accessible only to) a specific \verb`group`.

An \verb`interface_port` attribute lets Rosetta know on which port the container will expose its interface, and the \verb`interface_protocol` sets the corresponding protocol (e.g. HTTP, SSH, VNC etc.).
The \verb`interface_transport` (defaulted to TCP/IP) can be used to cover non-standard scenarios (e.g. if using UDP).

Since as explained in Section \ref{sec:architecture}, the container interfaces are made accessible to the outside world, they need to be secured. For this to happen, Rosetta allows to setup a one-time password or token at task creation-time to be used for accessing the task interface afterwards. Task interfaces can get password-protected in two ways: by implementing a password-based authentication at task-level, or by delegating it to the  proxy service. In the first case, the container must be built to support this feature and must be registered on the platform with the extra \verb`supports_interface_auth` attribute set to \verb`True`. Rosetta can then forward the password or token to the container via an environment variable. Instead, if the container makes use of an HTTP-based interface, it can delegate its access control to the proxy service, and just expose a plain, unprotected interface over HTTP. In this case, Rosetta will setup the proxy service in order enforce user authentication when accessing the task interface, and encrypt it using SSL. Delegating the task authentication to the proxy service is the default method for HTTP-based interfaces, since it is far more secure than leaving the authentication to be implemented at task-level, as it will be discussed in Section~\ref{sec:security}.

In order to support container engines missing port mapping capabilities, Rosetta provides a mechanism to let containers receive instructions on which port to start their interface on. As already mentioned in Section \ref{sec:architecture}, while most industry-standard container engines can autonomously manage TCP port mapping between containers and their host to avoid conflicts with ports already allocated (either by another service, by another container or by another instance of the same container), some of them cannot (e.g. Singularity).
In this case, the Rosetta agent can provide a specific port to the container where to make its interface to listen on, which is chosen between the free ephemeral ports of the host and passed to the container via an environment variable. To let Rosetta (and the agent) know that a given container supports this mechanism, its extra attribute \verb`supports_custom_interface_port` must be set to \verb`True` (and the \verb`interface_port` attribute is then discarded).

Rosetta comes with a number of a base containers for GUI applications, generic remote desktops and Jupyter Notebooks which can be easily extended to suit several needs:
\begin{itemize}
    \item JupyterNotebook, the official Jupyter Notebook container extended to support custom interface ports;
    \item GUIApplication, a container built to run a single GUI application with no desktop environment;
    \item MinimalDesktop, a desktop environment based on Fluxbox where more than one application can be run in parallel;
    \item BasicDesktop, a desktop environment based on Xfce for tasks requiring common desktop features as a file manager and a terminal.
\end{itemize}

The GUIApplication and Desktop containers make use of KasmVNC, a web-based VNC client built on top of modified versions of TigerVNC and NoVNC which provides seamless clipboard sharing between the remote application or desktop and the user's local desktop environment, as well as supporting dynamic resolution changes in order to always fit the web browser window, that are essential features in the everyday use.

%%======================================
%%  Implementation - Computing
%%======================================
\subsection{Computing}
\label{subsec:computing}
Computing resources are divided in two main types: \textit{standalone} and \textit{clusters}. The first ones may or may not have a WMS in place, while the second ones always do. If a computing resource has no WMS, the task execution is synchronous, otherwise the execution is asynchronous and the tasks are queued.

The Django ORM model class used to represent  computing resources is named \verb`Computing`, and it includes a \verb+type+, a \verb`name` and a \verb`description` fields for identifying a specific computing resource within Rosetta. A set of attributes describe how to access it and to submit user tasks: the \verb`access_mode` specifies how the computing resource is accessed (i.e. over SSH, using a command line interface (CLI), or a set of APIs); the \verb`auth_mode` specifies how the platform gets authorized on the computing resource; the \verb`wms` specifies the WMS in place (or if there is none) and the \verb`container_engine`  specifies which container engines (and runtimes) are available. With respect to the \verb`container_engine`, if the WMS natively supports containerized workloads and there is no need of running tasks using a specific container engine or runtime, then it can be just set to the value ``\verb+internal+''.

Some example combinations of these attributes are reported in Table \ref{table:computing}, where each row corresponds to a physical computing resource.
% Slurm
The first row represents a classic HPC cluster using Slurm as WMS and Singularity as container engine, and requiring an accredited cluster user to submit tasks over SSH using the Slurm command line interface.
The second row represents the same cluster but supporting, besides Singularty, also the Docker engine with both runC and Kata runtimes, in order to allow Rosetta (or its users) to chose the best one for a given task.
The third row represents yet the same cluster but accessed over Slurm REST APIs using JSON web tokens (JWT) for authentication.
% Standalone
The fourth and fifth rows represent instead standalone computing resources, using the Docker container engine, and accessed using SSH as a standard user for the fourth and the Docker REST APIs with a platform certificate for the fifth.
% Kubernetes
The sixth, seventh and eight rows all use computing resources managed with Kubernetes, and in the eight row the container runtimes available within Kubernetes are explicitly stated.
% Fargate
The last row is instead an example using Fargate, an hosted container execution service from Amazon Web Services (AWS) built on top of their Elastic Container Service (ECS), and accessed using its proprietary APIs.

When deploying Rosetta on a commercial cloud infrastructure as AWS or Google Cloud Platform (GCP), there are two options. The first one is to treat such infrastructures as transparent, and simply use standard (i.e. not proprietary) access modes as SSH, Slurm, or Kubernetes. In this case there is no difference between using Rosetta with computing resources deployed on premises or on such commercial cloud systems. The second option is to instead integrate at a deeper level, using AWS or GCP proprietary APIs and/or clients to automatically start new virtual machines upon request, or to use some of their native scheduling systems, as the last example of table \ref{table:computing}. 

The implementation work to support all of the combinations of access and authentication modes, container engines and WMSs is still ongoing, as we privileged SSH and Slurm since they fit well in the application scenarios we encountered so far. However, we wanted to lie down a general framework in order to easily expand the platform in future.

\begin{table*}[htpb]
\begin{center}
{\small %
\begin{tabular}{ |c|c|c|c|c|} 
\hline
% Computing resource
&\verb`access_mode` & \verb`auth_mode` & \verb`wms` & \verb`container_engines` \\
\hline
% Slurm
Computing resource \#1 & SSH+CLI & user keys & Slurm & Singularity \\
Computing resource \#2 & SSH+CLI & user keys & Slurm & Docker[runC,Kata],Singularity \\
Computing resource \#3 & API & JWT & Slurm & Docker,Singularity \\
% Standalone
Computing resource \#4 & SSH+CLI & user keys & none & Docker \\
Computing resource \#5 & API & platform cert. & none & Docker \\
% Kubernetes
Computing resource \#6 & CLI & platform cert. & Kubernetes & internal \\
Computing resource \#7 & SSH+CLI  & platform keys & Kubernetes & internal \\
Computing resource \#8 & API & platform cert. & Kubernetes & internal[runC,Kata] \\
% Fargate
Computing resource \#9 & API & platform cert. & Fargate & internal \\

\hline
\end{tabular}
}

\caption{Examples of various combinations of computing resource attributes. In order to schedule containerized workloads on a given computing resource, Rosetta needs to know how to access it (\texttt{access\symbol{95}mode}), how to get authorized (\texttt{auth\symbol{95}mode}), if and what WMS to use (\texttt{wms}), and which container engines are available (\texttt{container\symbol{95}engines}), possibly with their runtimes.}
\label{table:computing}
\end{center}
\end{table*}

The \verb`Computing` model describes the computing resource architectures as well, and in particular the \verb`arch` attribute defines the native architecture (e.g. amd64, arm64/v8), the \verb`supported_archs` attribute lists extra supported architectures (e.g. 386 on amd64 architectures) and the \verb`emulated_archs` attribute lists the architectures that can be emulated.

Computing resources can be also assigned to a specific group of users, using the \verb`group` attribute which, if set, restricts access to the group members only, and the  \verb`conf` attribute can be used to store some computing resource-specific configurations (e.g. the host of the computing resource). Lastly, the \verb+Computing+ ORM model implements an additional \verb`manager` property which provides common functionalities for accessing and operating on the real computing resource, as submitting and stopping tasks, viewing their logs, and executing generic commands. This property is implemented as a Python function which upon invocation instantiates and returns an object sub-classing the \verb`ComputingManager` class, based on the computing resource \verb`type`, \verb`access_mode`, \verb`auth_mode` and \verb`wms` attributes.

Computing resources which are accessed using SSH can be accessed both as a standard user (using its account on the computing resource) or using a superuser (e.g. a ``platform'' user), depending on the deployment requirements.

In order to access using a standard user on the computing resource, Rosetta generates a dedicated private/public key pair, the public key of which is required to be added on the computing resource account by the user. To instead access using a ``platform'' superuser (and thus using the same user for orchestrating all of the user tasks), a dedicated account and key pairs are required to be setup both on the computing resource and within Rosetta.

Accessing computing resources using SSH requires no integration with the existent infrastructure at all, provided that standard SSH is available and a container engine is installed. For this reason, it perfectly fits our requirement of operating on HPC clusters and data-intensive systems where more complex integrations are hard to achieve.

%%======================================
%%  Implementation - Storage
%%======================================
\subsection{Storage}

Storage functionalities provide a way of defining, mounting and browsing data storages. A \verb+Storage+ is defined by a set of attributes, which include a \verb+name+, a \verb+type+, an \verb+auth_mode+ and an \verb+access_mode+. 
If a storage is attached to a computing resource, then the \verb+computing+ attribute can be set. In this case, if the storage and the computing resource share the same access mode, the \verb+access_through_computing+ option can be ticked so that Rosetta can just use the computing resource one. The \verb+group+ attribute, if set, specifies the set of users authorized to access the storage. The \verb+base_path+ attribute sets the internal path to the storage, and supports using two variables: the \verb+$USER+, which is substituted with the Rosetta internal user name, and the \verb+$SSH_USER+, which is substituted with the SSH username (if the access method is based on SSH). The \verb+bind_path+ sets instead where the storage is made accessible within the software containers. If a data storage is attached to a computing resource and its \verb+bind_path+ is set, it will be then made accessible from all of the containers running on that computing resource, under the location specified by the \verb+bind_path+.

For example, a storage mounted on the \verb+/data+ mount point of an SSH-based computing resource (and represented in Rosetta using \verb+generic_posix+ as type and \verb~SSH+CLI~ as access method) could have a \verb+base_path+ set to \verb+/data/users/$USER+ and a \verb+bind_path+ set to \verb+/storages/user_data+, in order to separate data belonging to different users at orchestration-level.

At the moment only POSIX file systems are supported, which must be mounted on the various computing resources and that are in turn exposed inside the containers using the standard binding mechanism offered by most container engines. Any filesystem that can be mounted as such (e.g using FUSE) is therefore automatically supported, as CephFS or Amazon S3.

We envision adding support for other storage types in future releases, as for example object storages, but in this case accessing the storage APIs is up to the application running inside the container, and Rosetta can only act as a file manager. How to provide access in a standardized way to non-POSIX file systems within containers is indeed still an open theme.

Storage functionalities also include a set of APIs to provide support for the file manager embedded in the Rosetta web-based UI, which is built on top of the Rich File Manager \footnote{https://github.com/psolom/RichFilemanager} open source project. These APIs implement common functionalities (as get, put, dir, rename etc.) to perform file management operations, the internal logic of which depends on the storage type, making it easy to expand them in the future.

%%======================================
%%  Implementation - Tasks
%%======================================
\subsection{Tasks}
\label{subsec:tasks}

Tasks are represented using an ORM model and a set of states (\emph{queued}, \emph{running} or \emph{stopped}). Tasks running on computing resources without a WMS are directly created in the running  state, while when a WMS is in place they are created in the queued state and set as running only when they get executed.
States are stored in the \verb+state+ attribute of the \verb+Task+  model, which also includes a \verb+name+ and the links with the software container and the computing resource executing the task, plus its options (the \verb+container+, \verb+computing+ and \verb+computing_options+ attributes, respectively).
A set of other attributes as the \verb+interface_ip+, \verb+interface_port+, \verb+tcp_tunnel_port+ and \verb+auth_token+ let Rosetta know how to instantiate the connection to the task (i.e. for setting up the tunnel and/or configuring the proxy service).

Once a task starts on a computing resource, its IP address and port are saved in the corresponding \verb+Task+ fields, and the task is marked as running. If the task was queued, an email is sent to the user with a link to the task, which is particularly useful to let users immediately know when their tasks are ready, thus preventing to waste computing time on shared systems. Task functionalities also include opening the TCP/IP tunnel to the task interface port and/or configuring the HTTP proxy service in order to provide access to the task interface.

One of the main components of the task management functionalities is the agent, which as introduced in Section \ref{sec:architecture} allows to seamlessly support both WMSs not natively supporting containerized workloads and container engines missing some key features. In other words, it makes all of the computing resources behave in the same way from a Rosetta prospective. The agent is implemented as a Python script which is served by the Rosetta web application and that can run both as a superuser and as a standard, unprivileged user. When it is required, Rosetta delivers a bootstrap script on the computing resource which pulls and executes the agent code. As soon as it gets executed, the agent calls back the Rosetta web application and communicates the IP address of its host. If the agent landed on a computing resource using a container engine missing the dynamic port mapping feature, then it also searches for an available ephemeral TCP/IP port and communicates it to the web application as well. Lastly, the agent sets up the environment for the user task container, and starts it.

%%======================================
%%  Implementation - Account
%%======================================
\subsection{Account}
\label{subsec:account}
Account and profile functionalities provide support for both local and external authentication services (e.g. Open ID connect). The accounts linking between local and external identities is based on the user email address, which is the standard approach in this kind of services.

Local and external authentication can co-exist at the same time, provided that if a user originally singed up using an external authentication service it will be then always required to log-in using that service. If allowing to register as local users or to entirely rely on external authentication is up to the administrators, and can be configured in the web application service.

Rosetta provides both user-based and group-based authorization, so that computing and storage resources, as well as software containers, can be made available to specific users or subsets of users only. 

The user profile also supports some user-based configuration parameters for accessing the computing resources (e.g. the computing resource username if using an SSH-based access mode with user keys). Other minor functionalities, as password recovery, login tokens and time zone settings are provided as well.

\section{Security}
\label{sec:security}

Security of computing systems and web applications is a wide chapter and an extensive discussion on the topic is beyond the scope of this article, however we wanted to mention the main issues and how we took them into account.

The first layer of security in Rosetta consists in using software containers for the user tasks. The base executable unit in Rosetta is indeed the container itself, meaning that users has no control outside of their containers at all: once a container is sent for execution and Rosetta handles all the orchestration, the user is dropped inside it and cannot escape. 

For this reason, even if a container gets compromised, all the other ones as well as the underlying host system does not get affected.
However, this statement is true in the measure of which the container engine can guarantee isolation and prevent privilege escalation. The Docker engine has an intrinsic issue with this respect, as it makes use of a daemon running with superuser privileges. Podman, a nearly drop-in replacement for Docker, runs instead in user-space and prevents this kind of issues by design, as well as Singularity. Other container engines as gVisor and Kata push security even further, providing respectively kernel and hardware virtualization.

Moreover, when Rosetta is integrated on computing resources using SSH-based access, the administrators can opt for revoking direct SSH user access on them, leaving Rosetta - and its containerized tasks - the only access point, thus greatly improving overall security.

With respect to potential malicious software, the first line of defense usually takes place in the container registry. Docker Hub, for example, has a built-in security scanning system, and there are a number of free and open source scanners that can be used for on-premise container registries as Klar/Clair\footnote{\url{https://github.com/optiopay/klar}}. 

Scanning for malicious software can also be done when executing task containers\footnote{\url{https://docs.docker.com/engine/scan/}}, but not all container engines support this feature. Allowing only containers coming from registries which run security scanning, or to implement these checks along the building pipeline could be the best approach to protect against malicious software in container images \citep{brady2020docker}.

For what concerns software packages that can be installed at runtime inside the containers, Rosetta does not do any checking as it would be technically very hard if not even impossible. This is a common issue when giving users the freedom to download and execute code, including on commercial platforms as Google Colab and Kaggle. Even restricting user permissions would not prevent such issue, given that these packages can be always just downloaded and executed from a different location (e.g. a temporary folder). Having users to download and execute malicious software by mistake is therefore something very hard to prevent, and that has no simple mitigation approach unless relying on classic antivirus software which should run inside the containers.

As introduced in Section \ref{sec:architecture}, since Rosetta user task interfaces are made accessible to the outside world, they are required to be secured, both in term of access control and connection encryption. With this respect, it is necessary to make a distinction between HTTP-based and generic task interfaces. HTTP-based task interfaces can rely on the authentication and SSL encryption provided by the proxy service, and can therefore just use a plain HTTP protocol.
Generic task interfaces (e.g. a VNC or X server) are instead required to be secured at task-level, and it is responsibility of the task container to enforce it. As explained in subsection \ref{subsec:containers}, access control setup is in this case achieved by forwarding to the task a one-time password set by the user at task creation-time, which is then to be used by the container interface to authenticate the user. Encryption has to be setup at task-level too, and can be provided in first instance using self-signed certificates, or implementing more complex solutions as dynamic certificates provisioning.

An important detail in the task security context is that Rosetta makes a strong distinction between standard and power users, through a status switch in their profile. By default, only the latter can setup custom software containers using generic task interface protocols other than the HTTP, since handling security at task level (which is always required in this case) is error-prone and must be treated carefully. Standard users can therefore add and use custom software containers for their tasks on the platform only if using an HTTP-based interface, which is in turn forced to be secured by the proxy service.

For what concerns the tunnel from the web application service to the tasks, this is protocol-agnostic (above the TCP/IP transport layer) and is either accomplished by a direct connection on a private and dedicated network (e.g. if using Kubernets) or using an SSH-based TCP/IP tunnel using users' public/private keys, as explained in Section \ref{sec:architecture}, and thus assumed safe.

In terms of web security, we considered potential security risks originating from  cross-site request forgery (CSRF), cross-origin resource sharing (CORS), cross-site scripting (XSS), and similar attacks.
The same origin policy (SOP) of modern web browsers is already a strong mitigation for these attacks, and all the platform web pages and APIs (with a few exceptions for internal functionalities) uses Django's built-in CSRF token protection mechanism.
However, the SOP policy has limitations \citep{schwenk2017same,chen2018we}, in particular in our scenario where users can run custom (and possibly malicious) JavaScript code from within the platform, either using the Jupyter Notebooks or by other means (e.g. by setting up a task serving a web page).

We therefore focused on isolating user tasks from the rest of the platform even on the web browser side. Using the same domain for both the platform and the user tasks (e.g. \url{https://rosetta.platform/tasks/1} is indeed definitely not a viable solution as it does not allow to enforce the SOP policy at all. Also using dedicated subdomains (e.g. \url{https://task1.rosetta.platform}) has several issues, in particular involving the use of cookies \citep{zalewski2012tangled, zalewski2009browser, squarcina2021can}.

The secure-by-design, safe solution is to serve user tasks from a \textit{separate} domain (e.g. \url{rosetta-tasks.platform}). Then, each task can have its own subdomain (as \url{https://task1.rosetta-tasks.platform}) and stay separated form the main platform domain. However, handling and securing subdomains like this requires wildcard DNS services and SSL certificates, which for many institutional domains are not available \citep{jp_security}, including ours. For this reason, in Rosetta we opted for an intermediate solution: we serve user tasks from a separate domain (e.g. \url{rosetta-tasks.platform}) assigning each of them to a different port, under the same SSL certificate. In this way, the URL to reach the task number 1 at \url{https://rosetta-tasks.platform:7001} can be secured by the same SSL certificate covering the URL for task number 2 at \url{https://rosetta-tasks.platform:7002}, but are treated as different origins by web browsers.
SSL certificates are indeed port-agnostic, while the SOP (which basically involves the triplet protocol, host and port for defining the origin) it is not, thus enabling web browsers to enforce it between the task 1 and 2, and in general securing all of the users tasks against each others.
While this approach might lead to some issues with institutional firewalls blocking external port access beyond the standard 80 and 443 ports, we found it to be the right compromise in our environment. Moreover, switching to serving each task from its own subdomain is just a matter of a quick change in the Rosetta proxy service configuration.

%%======================================
%% The Rosetta science platform
%%======================================
\section{User Experience}
\label{sec:rosetta}

\begin{figure}[H]
\centering\includegraphics[scale=0.1,cfbox=gray 0.01ex 0ex]{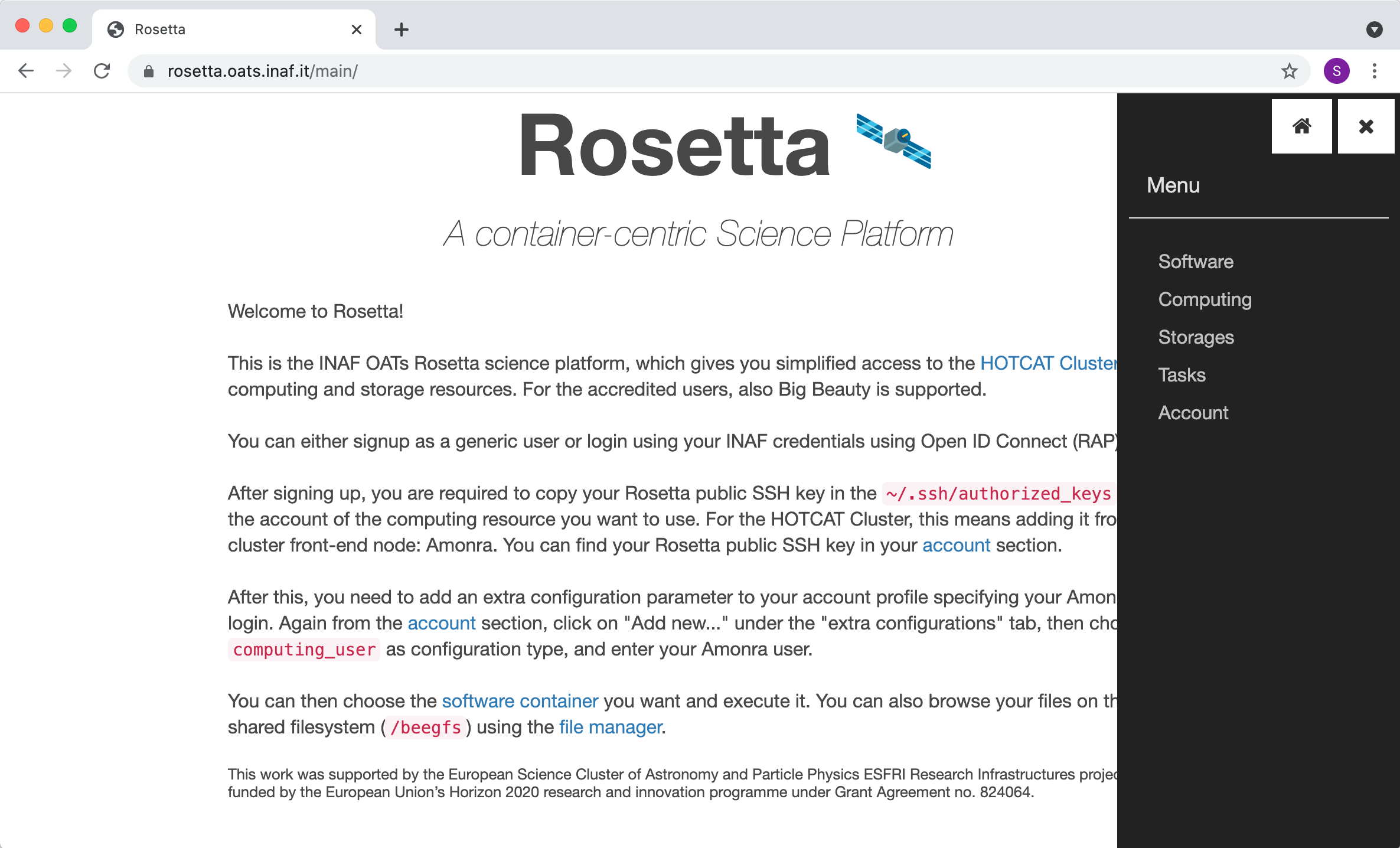}
\caption{The Rosetta science platform main page and menu.}
\label{fig:main}
\end{figure}

From a user prospective, Rosetta presents itself as a web application with a web-based user interface (UI) that is shown upon user login in Figure \ref{fig:main}.
The UI, following the architecture presented in section \ref{sec:architecture}, is organised in five main areas: the \emph{Software} section, where to browse for the software containers available on the platform or to add custom ones;
the \emph{Computing} section, where to view the available computing resources;
the \emph{Storage} section, which provides a file manager for the various data storages;
the \emph{Tasks} dashboard, where to manage and interact with the user tasks, including connecting with them and viewing their logs; 
and the \emph{Account} pages, where to configure or modify user credentials and access keys.\\

\begin{figure}[htpb]
\centering\includegraphics[scale=0.1,cfbox=gray 0.01ex 0ex]{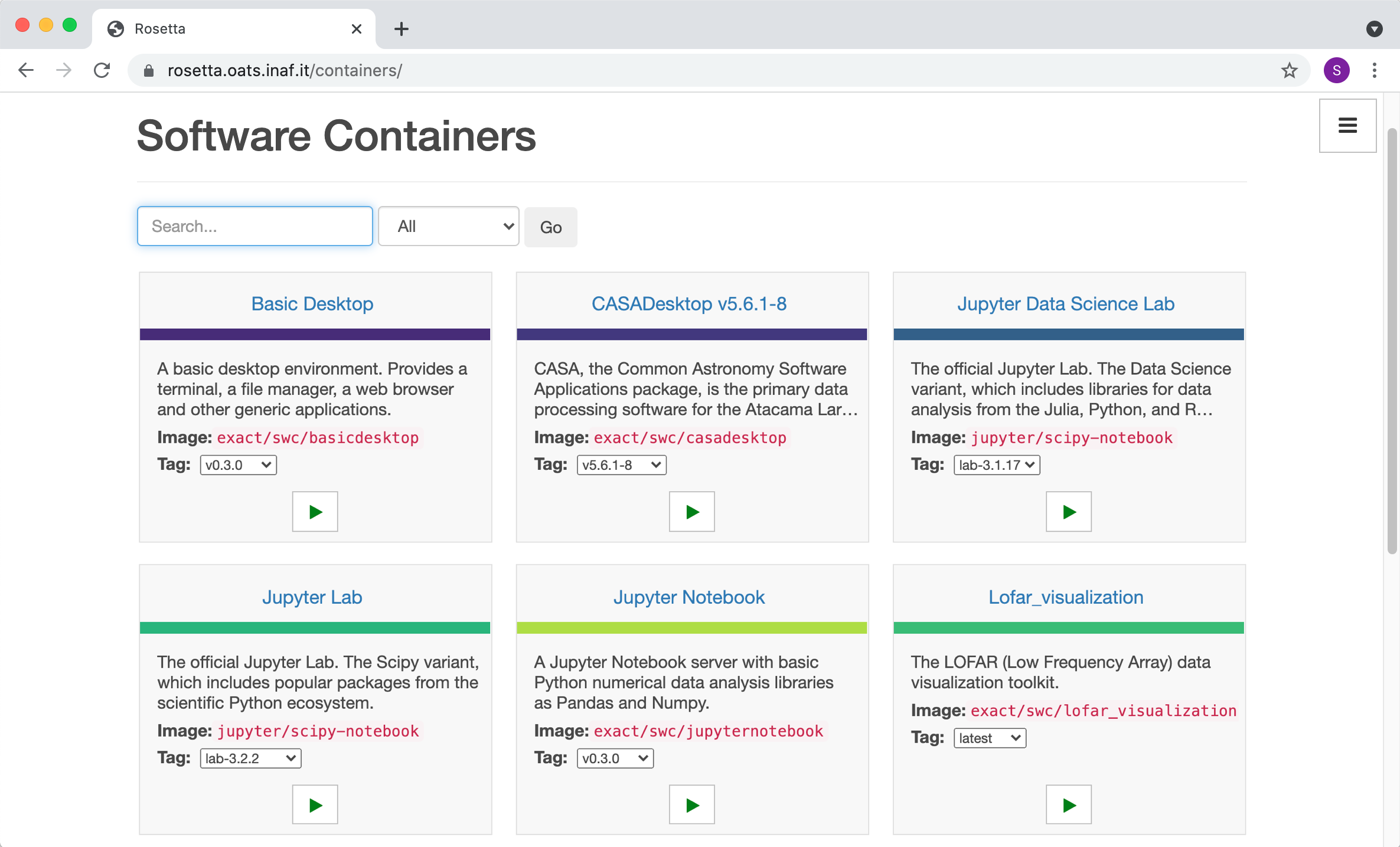}
\caption{Software containers list. For each software entry, a brief description is provided, together with the container image name  and a menu from which to select a specific version. The "play" button will start a new task with the given software.}
\label{fig:platform_containers}
\end{figure}

\begin{figure}[htpb]
\centering\includegraphics[scale=0.1,cfbox=gray 0.01ex 0ex]{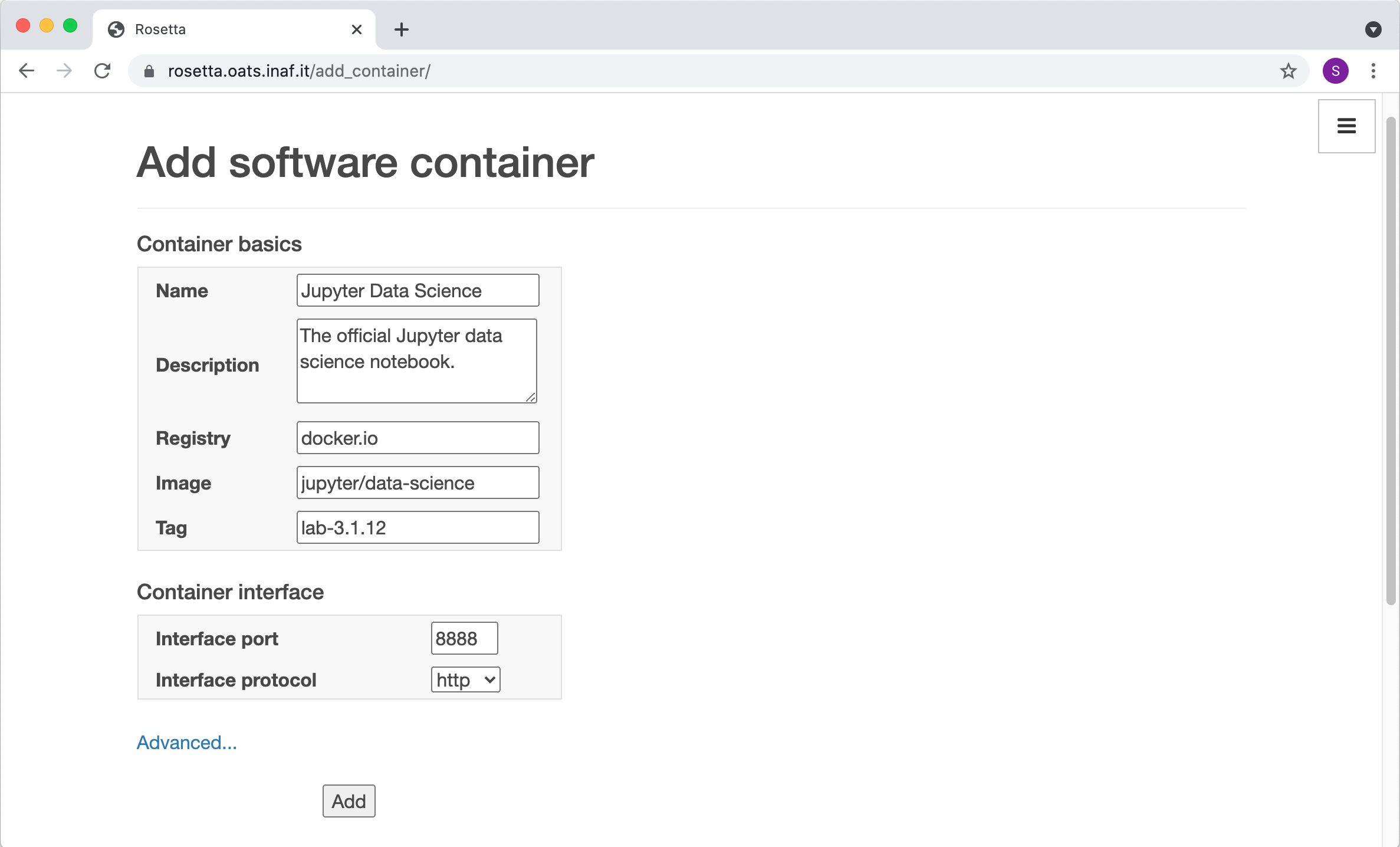}
\caption{Adding a new software container. Besides a name and a brief description, the key fields are the container registry, image and tag, plus the port and protocol of its interface. }
\label{fig:platform_add_container}
\end{figure}

To run a typical analysis task, the user first accesses the Software section (Figure \ref{fig:platform_containers}) in order to choose (or add) the desired software container. If adding a new software container, the user has to set its registry, image name, tag, the container interface port and protocol, plus some optional advanced attributes (Figure \ref{fig:platform_add_container}). The new container will then be listed together with the other ones so that the can be chosen for execution. 

\begin{figure}[htpb]
\centering\includegraphics[scale=0.1,cfbox=gray 0.01ex 0pt]{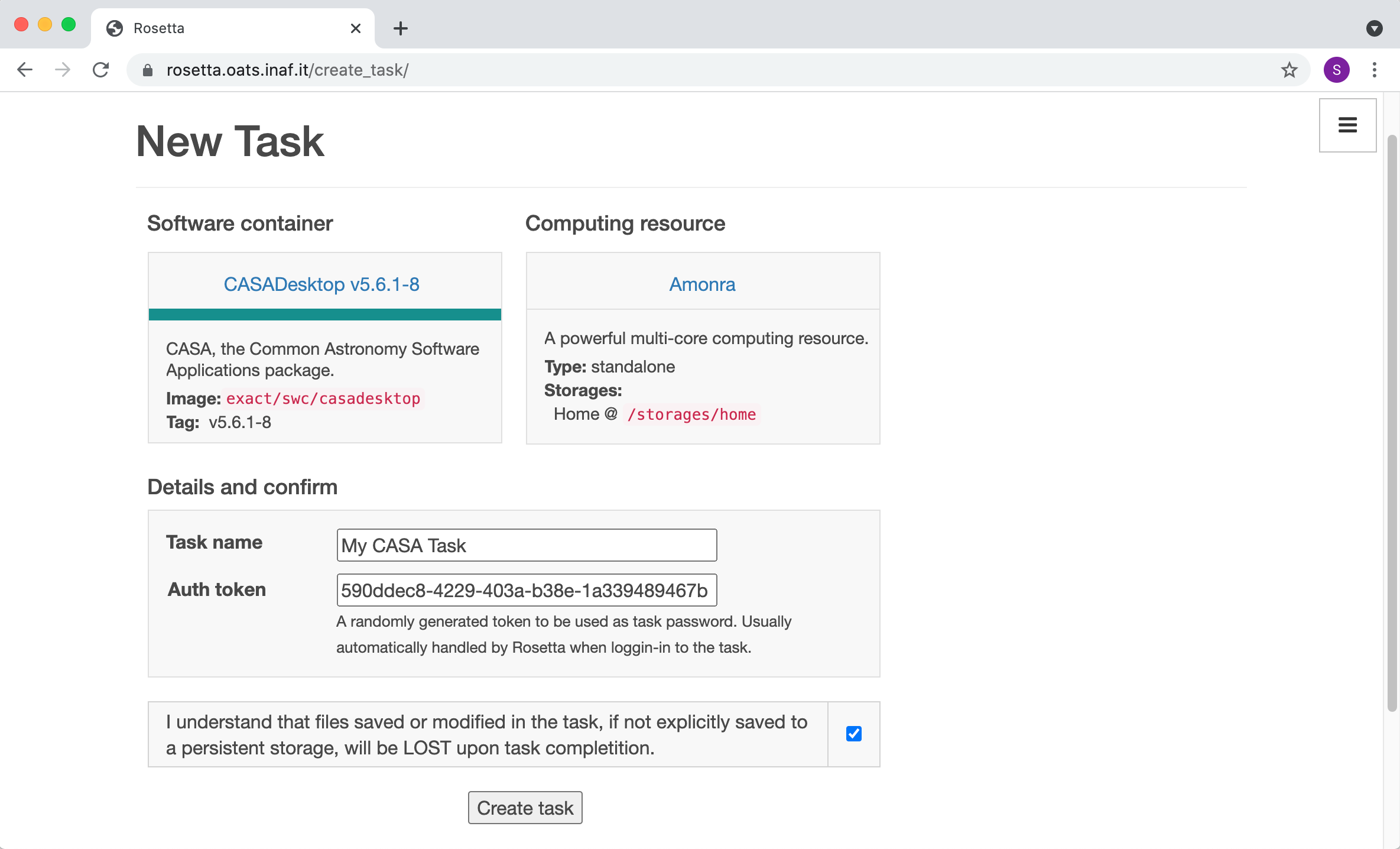}
\caption{Last step of new task creation, after selecting the software container and a computing resource. The interface asks to enter a task name and possibly other task parameters as the required number of CPUs, memory, or queue name.}
\label{fig:platform_new_task}
\end{figure}

Once the software container is chosen, the user hits the ``play" button to create a new task. The platform will then ask the user on which computing resource to run the task, and to set a task name. A one-time password token is also generated, which is usually automatically handled by Rosetta and not required to be entered manually when connecting to the task (Figure \ref{fig:platform_new_task}). For some computing resources, extra options as the queue or partition name, CPU cores and memory requirements can be set as well. The task is then created and submitted.

\begin{figure}[htpb]
\centering\includegraphics[scale=0.1,cfbox=gray 0.01ex 0pt]{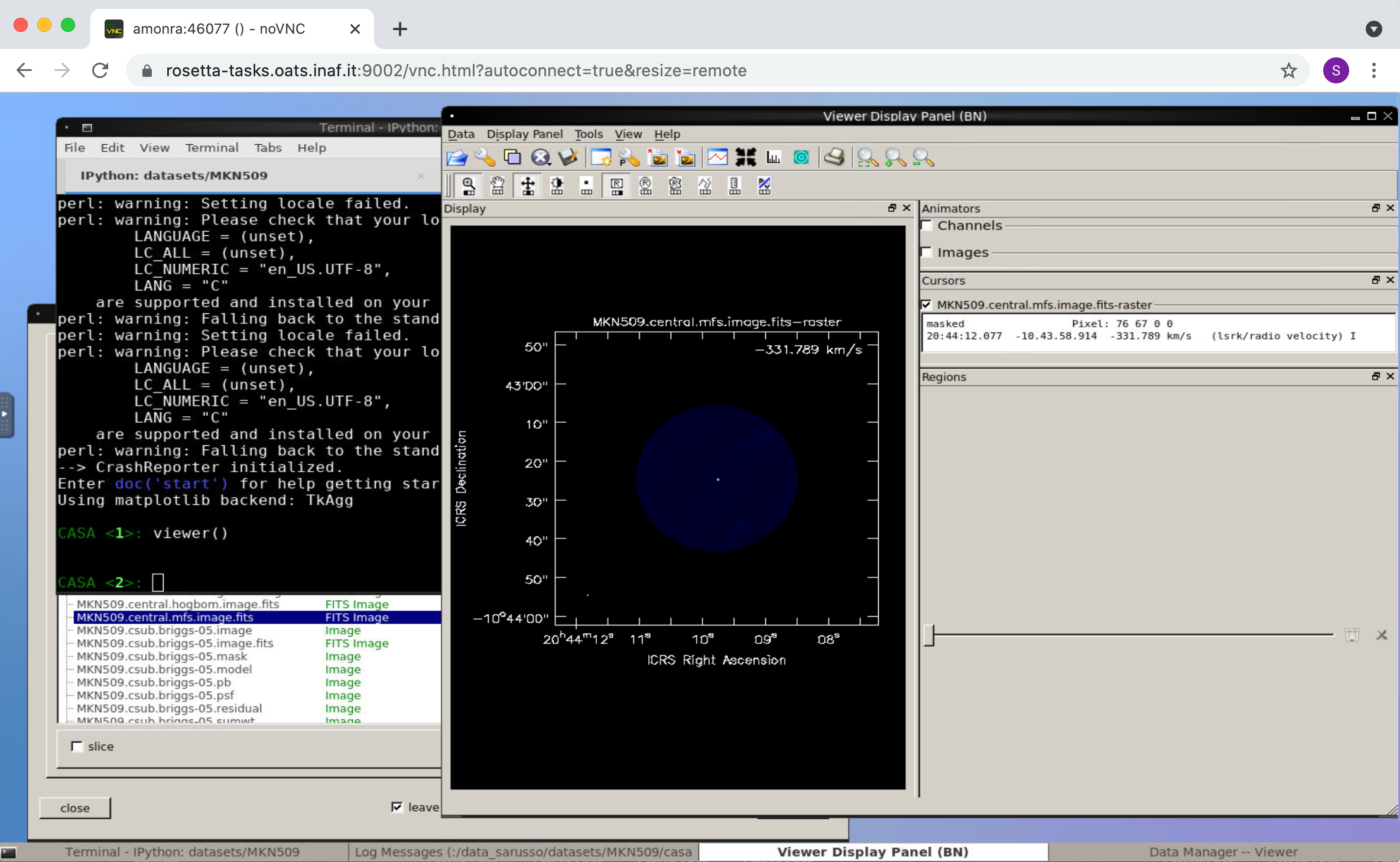}
\caption{A Rosetta user task running a GUI application from the CASA suite, in a remote desktop environment. The remote desktop server is web-based, and supports dynamic resolution changes and seamless clipboards sharing with the client, allowing for a smooth user experience.}
\label{fig:platform_task_CASA}
\end{figure}

\begin{figure}[htbp]
\centering\includegraphics[scale=0.11,cfbox=gray 0.01ex 0pt]{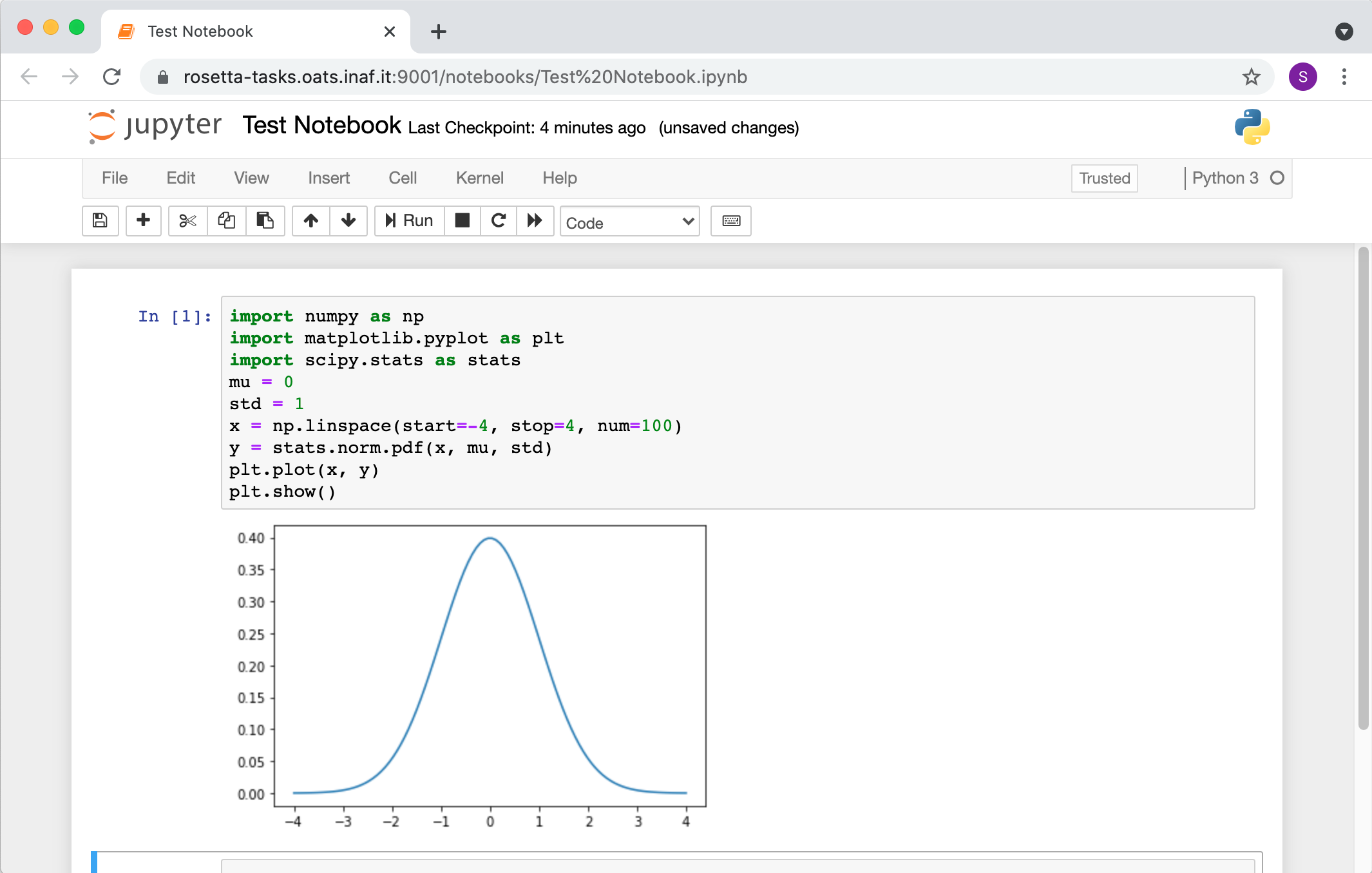}
\caption{A Rosetta user tasks running a Jupyter Notebook, displaying a plot using Numpy and Matplotlib. The authentication for the Notebook server is handled by the Rosetta proxy service, which also secures the connection over SSL.}
\label{fig:platform_task_Jupyter}
\end{figure}

As soon as the task is starting up on the computing resource, a ``connect'' button in the task dashboard becomes active. At this point, the user can connect to the task with just one click: Rosetta will automatically handle all the tunneling required to reach the task on the computing resource where it is running, and drop the user inside it
(Figures \ref{fig:platform_task_CASA} and \ref{fig:platform_task_Jupyter})
%and \ref{fig:platform_task_XCalc})

\begin{figure}[htpb]
\vspace{2.5mm}
\hspace*{0.08in}
\includegraphics[scale=0.1,cfbox=gray 0.01ex 0pt]{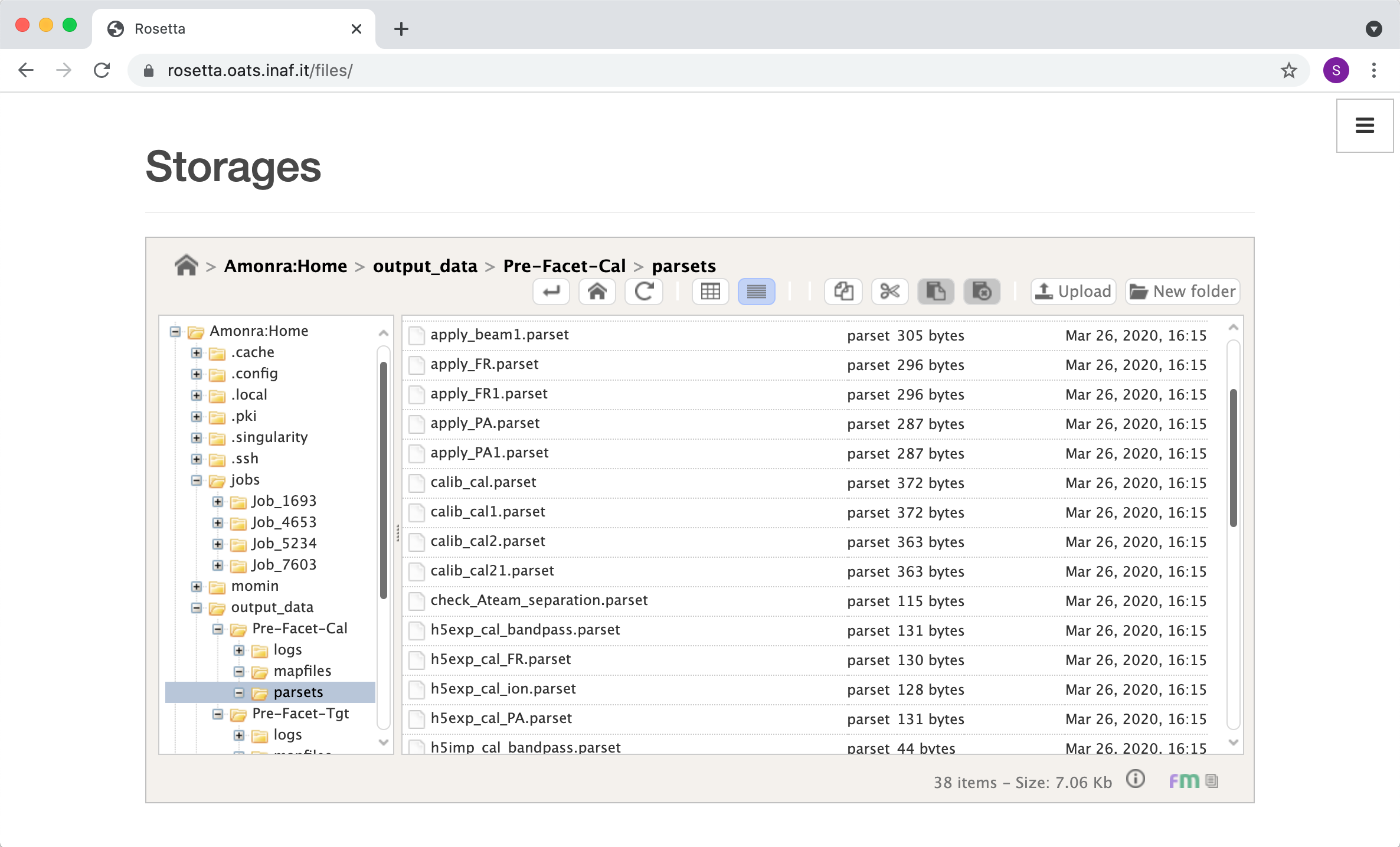}
\caption{The Rosetta built-in file manager, which allows for browsing data storages and to upload or download data files. While not suitable for large datasets, it is an effective tool for lighter ones as well as analysis scripts, plots and results.
\\}
\label{fig:platform_storages}
\end{figure}

Users can transfer files to and from the data storages (and thus the tasks) using the built-in file manager (Figure \ref{fig:platform_storages}), which is an effective solution for light datasets, analysis scripts, plots and results. Larger data sets are instead supposed to be already located on a storage, either because the data repository is located on the storage itself (in a logic of bringing the computing close to the data) or because they have been previously staged using an external procedures.
\\
\\

%\begin{figure}[htbp]
%\centering\includegraphics[scale=0.11,cfbox=gray 0.01ex 0pt]{images/SSH.png}
%\caption{A Rosetta user tasks running an SSH server with X protocol forwarding, rendering the XCalc application directly in the user graphical environment.}
%\label{fig:platform_task_XCalc}
%\end{figure}

%%======================================
%%    Deployment and use cases
%%======================================

\section{Deployment and use cases}
\label{sec:usecases}

Rosetta is deployed in production at the computing center of INAF - Osservatorio Astronomico di Trieste \citep{2020ASPC..527..303B}, using an extranal, aggregated authentication system named RAP \citep{tinarelli2020authentication} and serving a set of different users with different software requirements. 

To support our user community, we offer a pre-defined portfolio of containerized applications that span from generic data analysis and exploration tools (as iPython, R and Julia) to specific Astronomy and Astrophysics codes. These include common astronomical data reduction software and pipelines as IRAF, CASA, DS9, Astropy, but also Cosmological simulation visualization and analysis tools, and project-specific applications and codes.  All of them are listed in the Software section of Rosetta and are accessible from the users' web browsers by running a task instance.

In the following we discuss more in detail four different use cases among the various projects we support:
\textit{the LOFAR pipelines},
\textit{the SKA data challenges},
\textit{the Astrocook quasar spectral analysis software},
and \textit{the HPC FPGA bitstream design}.

\subsection{The LOFAR pipelines}

The software collection for the LOFAR community consists in a set of tools and pipelines used to process LOFAR data, as the Prefactor and DDFacet data reduction codes \citep{tasse2018faceting}, for which we created a set of software containers.

A typical run of the LOFAR data processing pipelines holds for several days, and requires significant computing resources (in terms of RAM, CPUs and Storage) to process terabytes of data ($\sim$ 15TB). Several checks are necessary during a pipeline run to verify the status of the data processing and the convergence of the results.

In this context, we are using Rosetta to run the pipelines within a software container that provides both the pipelines themselves and visual tools to check the status of the processing phase. Rosetta tasks run on an HPC cluster managed using the Slurm WMS, which allocates a set of resources in terms of RAM and CPUs as requested by the scientists in the task creation phase. These tasks compete with other standard Slurm jobs running on the cluster, thus ensuring an optimized allocation of the available resources among all users.

Scientist running the pipelines in this mode are not required to interact with the Slurm WMS or to manually deploy any software on the cluster, instead they can just rely on Rosetta and update the containers with new software if necessary.

The container source codes are available online as part of the LOFAR Italian collaboration \footnote{https://www.ict.inaf.it/gitlab/lofarit/containers} and once built are registered to an INAF private container registry in order to account for both public and private codes as required by the different LOFAR Key Projects collaborations.

\subsection{The SKA data challenges}

INAF participated in the SKA Data Challenges\footnote{https://sdc2.astronomers.skatelescope.org/sdc2-challenge} as infrastructure provider. The purpose of these challenges is to allow the scientific community to get familiar with the data that SKA will produce, and to optimise their analyses for extracting scientific results from them.

The participants in the second SKA Data Challenge analysed a simulated dataset of 1 TB in size, in order to find and characterise the neutral hydrogen content of galaxies across a sky area of 20 square degrees. To process and visualize such a large dataset, it was necessary to use at least 512 GB of RAM, and INAF offered a computing infrastructure where such resources were available.

We used Rosetta to provide simplified access to this computing infrastructure (an HPC cluster managed using the Slurm WMS) and, as for the LOFAR pipelines use case, we provided a software container that provided all of the tools and applications necessary to complete the challenge (as CASA, CARTA, WSClean, Astropy and Sofia) in a desktop environment.

Most notably, users were able to ask for specific computing resource requirements when starting their analysis tasks (512 GB of RAM, in this case), and the cluster parallel file system used to store the dataset provided high I/O performance ($>$ 4 GB/s) and plenty of disk space, so that users could focus on the scientific aspects of the challenge and not worry about orchestration and performance issues.

\subsection{The Astrocook quasar spectral analysis software}
Astrocook\citep{cupani2020astrocook} is a quasar spectral analysis software built with the aim of providing many built-in recipes to process a spectrum. While this software is not necessarily resource-intensive in general, it can require quite relevant computing power in order to apply the various recipes.

Astroccok comes as a GUI application with some common and less common Python dependencies which are sometimes hard to install (as Astropy, StatsModels and wxPython) and it is a great example about how to use Rosetta in order to provide one-click access to a GUI application which might require some extra computing power.

\begin{figure}
\centering\includegraphics[scale=0.11,cfbox=gray 0.01ex 0pt]{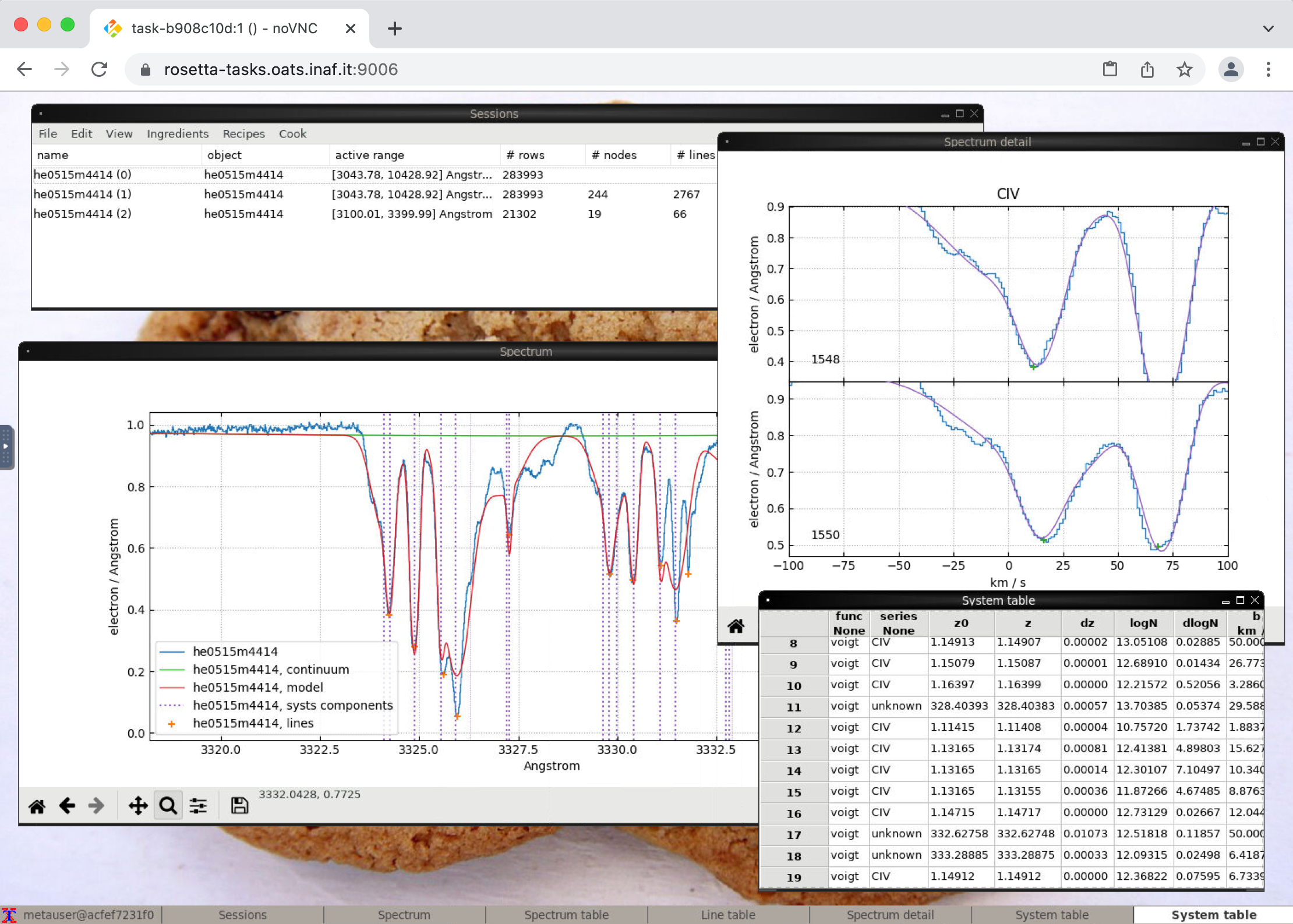}
\caption{The Astrocook quasar spectral analysis software running in a Rosetta task on a mid-sized computing resource. The "Spectrum" and "Spectrum detail" windows are the main components of Astrocook, while the "Sessions" and "System table" windows recap the analysis steps and parameters.}
\label{figure:astrocook}
\end{figure}

Figure \ref{figure:astrocook} shows Astrocook running in a Rosetta task on a mid-sized, standalone computing resource, and accessed using the web-based remote desktop interface.

\subsection{The HPC FPGA bitstream design}

Field Programmable Gate Arrays (FPGAs) can be used as accelerators in the context of physics simulations and scientific computing and they have been adopted as a low-energy acceleration devices for exascale testbeds.

One of these testbeds is ExaNeSt's (European Exascale System Interconnect and Storage) prototype \cite{KATEVENIS201858}, a liquid-cooled cluster composed by proprietary Quad-FPGA daughterboard computing nodes, interconnected with a custom network and equipped with a BeeGFS parallel filesystem.

To use this cluster it is necessary to re-engineer codes and algorithms \cite{9041710, 978-3-030-32520-6,computation8020034}: the substantial programming efforts required to program FPGAs using the standard approach based on Hardware Description Languages (HDLs), together with its subsequent weak code portability have long been the main challenges in using FPGA-enabled HPC clusters as the ExaNeSt's prototype.

However, thanks to the High Level Synthesis (HLS) approach, FPGAs can be programmed using high level languages, thus highly reducing the programming effort and and greatly improving portability. HLS tools use high level input languages as C, C++, OpenCL and SystemC which, after a process involving intermediate analysis, logic synthesis and algorithmic optimization, are translated into FPGA-compatible code as the so called ``bitstream" files.

This last step in particular requires a considerable amount of  resources: 128GB of RAM, extensive multi-threading support and 100 GB of hard disk space are the requirements for creating the bitstream files for the above mentioned FPGA-enabled HPC cluster. Moreover, from a user prospective, the design of an FPGA bitstream requires the interactive use of several GUI applications (as nearly all the HLS tools) and to let the software work for several hours.

Rosetta was adopted as the primary tool for programming INAF's FPGA cluster prototype, and suited very well the use case. Thanks to enabling access to persistent, web-based remote desktops with the required computing and storage resources, users were indeed capable of using HLS tools from their standard computing equipment, and to let them work for as many hours as needed, even if disconnecting and reconnecting the day after.
\\
\\
\\

%%======================================
%%        Discussion
%%======================================
\section{Discussion}
\label{sec:discussion}

In designing and implementing Rosetta we faced two main challenges: supporting custom software packages, libraries and environments, and integrating with computing resources not natively supporting containerized workloads.

We addressed the first challenge by developing a novel architecture based on framing user tasks as microservices. This allowed Rosetta to fully support custom software packages, libraries and environments (including GUI applications and remote desktops) and together with software containers allowed to ensure safe, consistent and reproducible code execution across different computing resources.

With respect to the second challenge, it has first to be noted that HPC clusters and data-intensive systems still rely on Linux users for a number of reasons, including accounting purposes and local permission management. This means that most of the containerisation solutions born in the IT industry, which assume to operate as a superuser, are in general not suitable. For this reason, the Singularity container engine was built to operate exclusively at user-level, and quickly become the standard in the HPC space.

However, Singularity is not designed to provide full isolation between the host system and the containers, and by default directories as the home folder, \verb+/tmp+, \verb+/proc+, \verb+/sys+, and \verb+/dev+ are all shared with the host, environment variables are exported as they are set on host, the PID namespace is not created from scratch, and the network and sockets are as well shared with the host. Also, the temporary file system provided by Singularity in order to make the container file system writable (which is required for some software) is a relatively weak solution, since it is stored in memory and often with a default size of 16MB, thus very easy to fill up.

We therefore had to address all these issues before being able to use Singularity as a container engine from Rosetta. In particular, we used a combination of command line flags (\verb`-cleanenv`, \verb`-containall`, \verb`-pid`) and ad-hoc runtime sandboxing for the key directories which require write access (as the user home), orchestrated by the agent. This step was key for the success of our approach and proved to remove nearly all the issues related to running Singularity containers on different computing systems.

Similarly, we had to work around a series of features lacking in WMSs not natively supporting containerized workloads (as Slurm), including container life cycle management itself, network virtualization and TCP/IP traffic routing between the containers, all solved using the agent as explained in the previous sections.

Once we were able to ensure a standardised behaviour of container engines and WMSs, we were able to make task execution uniform across different kinds of computing resources, providing the very same user experience. In this sense, Rosetta can be considered as an umbrella for a variety of computing resources, and can act as a sort of bridge in the transition towards software containers.

%%======================================
%%        Conclusions
%%======================================
\section{Conclusions and future work}
\label{sec:conclusions}

We presented Rosetta, a science platform for resource-intensive, interactive data analysis which runs user tasks as software containers. Its main characteristic lies in providing simplified access to remote computing and storage resources without restricting users to a set of pre-defined software packages, libraries and environments.

To achieve this goal, we developed a novel architecture based on framing user tasks as microservices - independent and self-contained units - which we implemented as software containers. This approach allowed us to fully support custom software packages, libraries and environments, including remote desktops and GUI applications besides standard web-based solutions as the Jupyter Notebooks. Moreover, adopting software containers allowed for safe, effective and reproducible code execution, and enabled us to let our users to add and use their own software containers on the platform.

We also took real-world deployment scenarios in mind, and designed Rosetta to easily integrate with existent computing resources, even where they lacked native support for containerized workloads. This proved to be particularly helpful for integrating with HPC clusters and data-intensive systems.

We successfully tested Rosetta for a number of use cases, including the LOFAR data reduction pipelines at INAF computing centers in the context of the ESCAPE project which funded this work, the SKA data challenges, and other minor use cases of our user community.

The benefits of seamlessly offloading data analysis tasks to a sort of ``virtual workstation", hosted on a computing system capable of providing CPUs, RAM and storage resources as per requests were immediately clear, removing constrains and speeding up the various activities.

Although astronomy and astrophysics remains its mainstay, Rosetta can virtually support any science and technology domain requiring resource-intensive, interactive data analysis, and it is currently being tested and evaluated in other institutions.

Future work include adding support for distributed workloads (e.g. MPI, Ray) and for computing resources with mixed architectures, developing a command line interface, integrating with data staging solutions and continuing the implementation efforts for integrating with current and new WMSs (e.g. Torque, Openshift, Rancher, Nomad, and more).

%%======================================
%%  Acknowledgements
%%======================================
\section{Acknowledgements}
\label{sec:ack}
This work was supported by the European Science Cluster of Astronomy and Particle Physics ESFRI Research Infrastructures project, funded by the European Union’s Horizon 2020 research and innovation programme under Grant Agreement no. 824064. We also acknowledge the computing center of INAF- Osservatorio Astronomico di Trieste, \citep{2020ASPC..527..303B,2020ASPC..527..307T}, for the availability of computing resources and support.

\bibliography{rosetta}

\end{document}